\documentclass[12pt]{iopart}
\usepackage{iopams}

\begin{document}

\title[Comment]
{Comment to the reply to "On the thermodynamics of inhomogeneous
perfect fluid mixtures"}
\author{Bartolom\'e Coll$^1$\
and Joan Josep Ferrando$^2$}

\address{$^1$\ Syst\`emes de r\'ef\'erence relativistes, SYRTE,
Obsevatoire de Paris-CNRS, \\75005 Paris, France.}

\address{$^2$\ Departament d'Astronomia i Astrof\'{\i}sica, Universitat
de Val\`encia,\\46100 Burjassot, Val\`encia, Spain.}
\ead{bartolome.coll@obspm.fr; joan.ferrando@uv.es}

\begin{abstract}
In spite of a recent reply by Quevedo and Z\'arate (gr-qc/0403096), 
their assertion that their thermodynamic scheme
for a perfect fluid binary mixture is incompatible with Szekeres
and Stephani families of universes, except those of FRW ones,
remains wrong.

\end{abstract}


In the paper ``Thermodynamic scheme of inhomogeneous perfect fluid
mixtures" \cite{zq}, Quevedo and Z\'arate introduce a
thermodynamic scheme for a binary mixture of relativistic perfect
fluids with non-vanishing entropy production, and analyze its
compatibility with the Szekeres and Stephani families of perfect
fluid solutions to Einstein Equations.

Their thermodynamic scheme is very clearly presented, the two
families of solutions are well known from long time, and in any
moment of their argumentation additional restrictions are made due
to physical considerations or to supposedly known variables. In
this {\em general} context fixed by them, they conclude that their
thermodynamic scheme is incompatible with {\em all} the
space-times of both families of solutions except the
Friedmann-Robertson-Walker ones.

This conclusion of incompatibility, repeatedly asserted in their
paper \cite{zq}, is wrong.

In our Comment "On the thermodynamics of inhomogeneous perfect
fluid mixtures" \cite{CFcommnet}, we proved the wrongness of their
assertion by {\em three independent} ways: \vspace{1.5mm}

i) giving the correct analysis of compatibility and proving that,
contrarily to their assertion, not only {\em all} the perfect
fluid space-times admit their scheme, but that every one of them
admits {\em many} different such schemes, \vspace{1.5mm}

ii) showing the errors in two of their crucial arguments that lead
them to their false conclusion and \vspace{1.5mm}

iii) giving simple counterexamples to their assertion.
\vspace{1.5mm}

Surprisingly enough, their Reply \cite{QZ} to our Comment: \vspace{1.5mm}

i') only presents vague comments about {\em imprecise} 'physical
meanings' and 'known variables', not referred to in their original
paper, which anyway are {\em strongly insufficient} to constitute
a set of forgotten or implicit conditions in \cite{zq} that could
be able to transform their wrong assertion into a correct one.
\vspace{1.5mm}

ii') does not consider the two errors in their crucial arguments
that lead them to their false conclusion, \vspace{1.5mm}

iii') does not takes into account the counterexamples to their
assertion. \vspace{1.5mm}

As a consequence of these three points (but only one of them
suffices!), {\em the mentioned assertion, even to the light of
their replay} \cite{QZ}{\em, remains clearly wrong}, as indeed may
be verified by any advanced student in the subject.

Our intention has not been, and is not, polemic. Essentially we
try to avoid other searches, particularly those non specialists
but interested in the subject, to lose their time (and money ?)
studying or assuming such a wrong result.

For this reason and the strong evidence of the errors in paper
\cite{zq}, from now on we shall not comment on any other vague
argumentation on this subject. \\[4mm]
{\bf NOTES} \vspace{1mm}

1. In their Reply \cite{QZ} it is asserted that the conclusion of
our Comment \cite{CFcommnet} is erroneous because based on an
erroneous assumption (namely, that the chemical potential and
fractional concentration of a mixture of perfect fluids are
unknown variables). This assertion is {\em impertinent} because
{\em hypotheses non f\'ecimus} in our Comment, at least neither
more nor less than them in their original paper \cite{zq}.
\vspace{1mm}

2. In their Reply \cite{QZ} it is  asserted that in our Comment
\cite{CFcommnet} we apply inappropriately the 'Pfaffian method' to
a 'non Pfaffian' form. This assertion is also {\em impertinent}
because the only result from J. F. Pfaff that we use is his local
decomposition theorem for {\em arbitrary} one-forms, later
recovered by G. Frobenius, G. Darboux and E. Cartan among others.
Although one may well be unaware of these historical details, our
description of the contents of the theorem ('local
decomposition'), our explicit expression (5) and the simplicity of
the context makes hard to understand that the word 'Pfaff' could
so perturb and so confuse the authors of the Reply \cite{QZ}. It is
obvious that, contrarily to their assertion, the (Pfaff)
decomposition theorem used in our Comment is not only {\em
completely} appropriate but is {\em the} appropriate one to be
applied. \vspace{1mm}

3. It remains, as it is evident, that the two families of
solutions in question cannot admit arbitrarily given binary
mixtures. The problem of finding all the (formally) admissible
binary mixtures, and that of finding, among them, those which may
be physically interesting for every particular application remain
open. Nevertheless, and as already stated in our Comment
\cite{CFcommnet} (last paragraph of Section 2), the richness of
the choice of Z\'arate and Quevedo's schemes stated in our
Proposition 2 locally guarantees the usual thermodynamic
inequalities (such as $T > 0,$ or $0 \leq c \leq 1$).

\end{document}